\documentclass[sigconf]{acmart}

\usepackage{subcaption}
\usepackage{listings}
\usepackage{xcolor}

\definecolor{codegreen}{rgb}{0,0.6,0}
\definecolor{codegray}{rgb}{0.5,0.5,0.5}
\definecolor{codepurple}{rgb}{0.58,0,0.82}
\definecolor{backcolour}{rgb}{0.95,0.95,0.92}
\lstdefinestyle{mystyle}{
    backgroundcolor=\color{backcolour},   
    commentstyle=\color{codegreen},
    keywordstyle=\color{magenta},
    numberstyle=\tiny\color{codegray},
    stringstyle=\color{codepurple},
    basicstyle=\ttfamily\footnotesize,
    breakatwhitespace=false,         
    breaklines=true,                 
    captionpos=b,                    
    keepspaces=true,                 
    numbers=left,                    
    numbersep=5pt,                  
    showspaces=false,                
    showstringspaces=false,
    showtabs=false,                  
    tabsize=2
}

\AtBeginDocument{%
  }

\renewcommand\footnotetextcopyrightpermission[1]{}
\begin{document}

\title{Scalable Packed Layouts for Vector-Length-Agnostic ML Code Generation}

\author[Beysel et al.]{Ege Beysel}
\email{beysel@roofline.ai}
\orcid{0000-0002-3176-4578}
\affiliation{%
  \institution{RooflineAI}
  \city{Cologne}
  \state{NRW}
  \country{Germany}
}

\author{Maximilian Bartel}
\email{bartel@roofline.ai}
\affiliation{%
  \institution{RooflineAI}
  \city{Cologne}
  \state{NRW}
  \country{Germany}
}

\author{Jan Moritz Joseph}
\email{joseph@roofline.ai}

\affiliation{%
 \institution{RooflineAI}
  \city{Cologne}
  \state{NRW}
  \country{Germany}
}

\affiliation{%
  \institution{RWTH Aachen University}
  \city{Aachen}
  \state{NRW}
  \country{Germany}
}



\begin{abstract}
Scalable vector instruction sets such as Arm SVE enable vector-length-agnostic (VLA) execution, allowing a single implementation to adapt across hardware with different vector lengths. However, they complicate compiler code generation, as tiling and data layout decisions can no longer be fixed at compile time.

We present an approach for enabling VLA code generation in an end-to-end ML compilation pipeline through vector-length-aware packed data layouts and corresponding compiler extensions. We integrate these mechanisms into MLIR/IREE and extend tiling, fusion, and vectorization to operate with scalable vector lengths.

Evaluated on real-world ML workloads on Arm CPUs, our approach generates SVE code that is competitive with, and often outperforms, existing NEON-based code generation within IREE, achieving up to $1.45\times$ speedup. We also outperform PyTorch ecosystem frameworks, including ExecuTorch, TorchInductor, and eager execution, demonstrating the effectiveness of scalable vectorization in a production compiler setting. A simulator-based study further shows that the generated code scales with increasing SVE vector length on compute-bound workloads, supporting performance portability across hardware configurations.
\end{abstract}



\maketitle
\pagestyle{plain}
\section{Introduction}

The increasing deployment of machine learning (ML) workloads across diverse environments beyond datacenters, particularly on edge and consumer devices, has increased the importance of efficient execution on a wide range of hardware platforms. In particular, CPUs remain a key deployment target due to their ubiquity, flexibility, and ability to support diverse workloads without requiring specialized hardware. At the same time, ML workloads must often be deployed across hardware generations and system configurations without per-device recompilation, despite differences such as vector length and microarchitectural details, making efficient and general compilation strategies essential.

Modern CPU architectures provide SIMD (Single Instruction, Multiple Data) extensions to exploit data-level parallelism. Widely deployed instruction sets such as Arm NEON use fixed vector lengths, enabling highly efficient implementations when software is tuned to a specific target. However, this model requires retuning across architectures with different vector lengths. More recent scalable vector instruction set architectures (ISAs), such as Arm's Scalable Vector Extension (SVE)~\cite{stephens2017arm}, introduce a vector-length-agnostic (VLA) programming model, where the vector length is implementation-defined within an ISA-specified range. This enables a single implementation to adapt across different hardware configurations while preserving performance portability, but also introduces new challenges for compiler design.

End-to-end ML compilers have emerged as an effective approach for bridging high-level frameworks and low-level hardware execution. Modern compiler stacks such as Intermediate Representation Execution Environment (IREE)~\cite{liu2022tinyiree} operate on whole computation graphs and apply progressive lowering from high-level tensor operations to low-level implementations. These systems are typically built on compiler infrastructures such as Multi-Level Intermediate Representation (MLIR)~\cite{mlir}, which provides the abstractions and transformation mechanisms required to represent and lower such workloads.

This end-to-end, MLIR-based compilation approach enables systematic transformations and generalization across different model architectures and target backends, in contrast to traditional per-operator kernel implementations. However, existing compiler pipelines are largely designed around fixed-length vectorization and do not directly account for the constraints and opportunities introduced by scalable vector architectures.

In this work, we present an approach for integrating scalable vector extensions into an end-to-end ML compilation pipeline through \textit{vector-length-parametric packed data layouts} and compiler transformations. Our design enables efficient lowering of tensor operations to scalable vector code while preserving portability across hardware with different vector lengths. By propagating packed layouts throughout the compilation pipeline and adapting tiling, fusion, and vectorization passes accordingly, we achieve a consistent mapping from high-level tensor programs to VLA implementations.

We evaluate our approach on a diverse set of real-world models, including transformer-based language models, vision models, and audio models, across multiple Arm-based platforms. Our results show that the generated SVE code is competitive with, and often outperforms, existing NEON-based code generation within IREE, achieving up to $1.45\times$ speedup despite identical vector lengths. Furthermore, we outperform production and research frameworks from the PyTorch~\cite{paszke2019pytorch, ansel2024pytorch} ecosystem, including ExecuTorch~\cite{executorch}, TorchInductor, and PyTorch's default eager execution mode, with speedups of up to $1.70\times$ (and up to $12.38\times$ on Pixel 9), $6.09\times$, and $3.67\times$, respectively, particularly on matrix multiplication--dominated workloads. Additionally, a controlled simulator-based study confirms that our generated code scales as expected with increasing SVE vector length, reaching up to $3.44\times$ speedup when moving from a 128-bit to a 512-bit SVE implementation on compute-bound workloads. These results demonstrate that scalable vectorization can be integrated into a production compiler stack while maintaining high performance and enabling portability across hardware generations.

The main contributions of this paper are as follows:
\begin{itemize}
    \item We introduce vector-length-parametric packed data layouts that enable efficient and portable execution of tensor operations on scalable vector architectures.
    
    \item We integrate support for scalable packed layouts into a production-grade end-to-end ML compilation pipeline based on MLIR and IREE, and extend core compiler transformations, including tiling, fusion, and vectorization, to operate correctly and efficiently on vector units with scalable lengths.
    
    \item We enable VLA code generation in this pipeline and demonstrate the approach on Arm SVE using a range of real-world ML workloads dominated by FP32 matrix multiplication, showing competitive and improved performance over existing systems on real hardware, as well as performance scaling with SVE vector length in a controlled simulation model.

    \item We provide an open-source implementation of our approach, including compiler extensions to IREE and LLVM, and are actively working towards upstreaming these contributions to the respective mainline projects~\cite{UPSTEAMLINK}.
    
\end{itemize}
The mechanisms developed in this work lay the groundwork for extending the approach beyond the evaluated setting of FP32 matrix multiplication on Arm SVE to other data types such as BF16 and INT8, to scalable vector ISAs such as Arm Scalable Matrix Extension (SME)~\cite{arm_sme} and RISC-V Vector (RVV)~\cite{rvv_spec}, and to other kernel-specific packed layouts.

The remainder of this paper is structured as follows. Sec.~\ref{sec:background} provides background on scalable vector architectures, packed matrix multiplication, and ML compiler infrastructures. Sec.~\ref{sec:related_work} reviews prior work in vector-length-agnostic code generation and ML compilation. Sec.~\ref{sec:methods} presents our approach, including scalable packed layouts and their integration into the compiler pipeline. Sec.~\ref{sec:results} evaluates our implementation across a range of workloads and compares it to existing systems. Finally, Sec.~\ref{sec:conclusion} concludes the paper.
\section{Background}
\label{sec:background}

This section introduces the hardware target (Arm SVE) followed by the relevant computational and compiler abstractions used in this work.

\subsection{Arm Scalable Vector Extension}

The Arm Scalable Vector Extension (SVE) is a SIMD extension for AArch64 designed to overcome the limitations of fixed-length vector ISAs such as NEON, particularly their lack of portability across processors with different vector lengths. Unlike NEON, which operates on 128-bit vectors, SVE introduces a VLA programming model, where the hardware vector length is a runtime constant chosen by the hardware implementation, with the SVE specification allowing vector lengths from 128 to 2048 bits in powers of two. This allows a single binary to execute across processors with different vector lengths without recompilation. This flexibility, however, introduces challenges for code generation, as key parameters such as vector length are not known at compile time and must be handled symbolically or deferred to runtime. This complicates optimizations such as loop tiling and vectorization, which typically rely on statically known tile sizes and data layouts.

SVE provides 32 scalable vector registers (\texttt{Z} registers) and introduces per-lane predication through dedicated predicate registers (\texttt{P} registers). Predication enables efficient handling of non-multiple-of-vector-length problem sizes by masking inactive lanes, avoiding scalar remainder loops. Additional features such as gather/scatter memory operations and vector-friendly loop constructs further improve programmability for data-parallel workloads.

SVE2 extends this model to a broader set of application domains, while Arm SME introduces a streaming mode with a potentially larger streaming vector length (SVL), as well as new architectural components such as registers for 2-dimensional tiles and dedicated outer-product units for accelerating matrix computations. Although this work focuses on SVE, the principles of VLA code generation extend naturally to other scalable vector ISAs, such as RVV and Arm SME.

\subsection{Matrix Multiplication and Data-Tiled Layouts}

Matrix multiplication is a central operation in machine learning workloads, particularly in transformer-based models. Efficient implementations~\cite{goto2008anatomy, smith2014anatomy} rely on data tiling, often realized through operand packing, a form of memory layout transformation that reorganizes data in memory to improve cache locality and SIMD utilization.

In this context, packing transforms input matrices into layouts that match the access patterns expected by optimized compute kernels, which are themselves designed around the capabilities of the underlying hardware. Instead of operating on row-major matrices directly, the computation is performed on smaller tiles stored contiguously in memory. This reduces cache misses and enables efficient vectorized execution in the innermost loops.

Conventional code generation approaches use \emph{static packed layouts}, where tile sizes are fixed at compile time, typically based on the target vector length and chosen kernel configuration. In contrast, this work introduces \emph{scalable packed layouts}, where tile dimensions are parameterized by the vector length of the hardware implementation, which is not known at compile time. This enables the generated code to remain portable across different hardware implementations without requiring retuning or recompilation.

\subsection{Machine Learning Compilers}

ML compilers translate high-level models from frameworks such as PyTorch or TensorFlow into efficient executables for diverse hardware targets. Unlike traditional compilers, ML compilers operate on computational graphs and apply domain-specific optimizations such as operator fusion, data layout transformations, tiling, and parallelization.

A key advantage of ML compilers is their ability to perform \emph{whole-graph optimization}, enabling transformations that span multiple operators and better align computation with hardware characteristics. This contrasts with library-based approaches, which rely on pre-optimized kernels and typically optimize operators in isolation.

ML compilers also play a central role in targeting diverse hardware backends, including CPUs, GPUs, and specialized accelerators. This abstraction enables them to support architectural diversity and evolving instruction set features, while still generating code that exploits the capabilities of each target, including features such as scalable vector extensions. As a result, they enable performance portability across devices and reduce the need for manually tuned, hardware-specific kernels.

\subsection{MLIR and IREE}

MLIR provides an extensible compiler infrastructure that supports multiple levels of abstraction within a unified framework. Instead of relying on a single fixed IR, MLIR allows domain-specific dialects to represent computations at different stages of lowering, from high-level operations down to hardware-specific code. This design is particularly well-suited for ML workloads, where transformations must bridge large semantic gaps between frameworks and hardware.

IREE is an MLIR-based machine learning compiler and runtime that targets a wide range of backends, including CPUs, GPUs, and accelerators. It follows a compiler-driven approach, progressively lowering models from high-level representations into efficient device-specific code. Unlike library-based systems, IREE does not rely on hand-written kernels, but instead generates code through systematic transformations and scheduling. 

In a typical compilation flow, models are imported into MLIR (e.g., via \texttt{torch-mlir}~\cite{torchmlir}) and lowered through a sequence of dialects, including \texttt{linalg} for structured tensor computations and \texttt{vector} for explicit vectorization. Target-specific dialects, such as \texttt{arm\_sve}, are then used to represent scalable vector operations before lowering to the \texttt{llvm} dialect, LLVM IR, and finally target binaries (e.g., AArch64). This progressive lowering enables architecture-specific optimizations while maintaining a high-level representation of the computation.

In this work, we extend IREE’s CPU backend to support scalable data-tiled layouts for SVE, enabling VLA data layout and code generation within an end-to-end ML compilation pipeline.
\section{Related Work}
\label{sec:related_work}

Prior work on VLA code generation can be broadly grouped into compiler support and autovectorization studies, kernel- and library-level approaches, and end-to-end compiler integrations for ML inference workloads. These works highlight both the promise of scalable vector ISAs such as Arm SVE and RVV and the challenges of generating efficient code for them.

At a fundamental level, several studies analyze the performance implications of VLA programming models. Pohl et al.~\cite{pohl2019performance}, and Poenaru and McIntosh-Smith~\cite{poenaru2020evaluating} evaluate the effectiveness of vector-length-agnostic ISAs and show that, while VLA enables portability across hardware implementations, achieving high performance depends on how well compilers and runtimes adapt to the underlying vector length. Subsequent work on RVV compiler support further highlights gaps in autovectorization and code generation compared to hand-optimized implementations~\cite{adit2022performance, carpentieri2025performance, lai2025risc}. These studies motivate the need for improved compilation strategies for scalable vector architectures.

A second line of work focuses on generating efficient kernels for VLA ISAs. For example, Igual et al.~\cite{igual2023automatic} propose automatic generation of matrix multiplication microkernels for RVV, targeting performance portability across implementations. Their results show that, while VLA enables functional portability, achieving peak performance still requires fine-grained, implementation-specific tuning due to microarchitectural differences. In contrast, our work encodes vector-length agnosticism directly in the data layout by introducing packed layouts whose tile dimensions adapt to the implementation-defined vector length, enabling high-performance code generation without per-implementation retuning. While highly specialized kernels may still be required to reach peak performance on a given microarchitecture, our approach achieves competitive performance across full ML inference workloads within an end-to-end compilation pipeline.

More recently, end-to-end compiler approaches have begun to expose VLA execution to machine learning workloads. Work from Kalda and Hutton~\cite{kalda2024sve_sme_ml} on integrating Arm SVE and SME into Apache TVM~\cite{chen2018tvm} enables VLA code generation and introduces schedules for selected operators such as convolution and matrix multiplication. However, these approaches primarily express vector-length dependence through loop transformations and tensor intrinsics, and do not explicitly model data-tiling or packing as first-class abstractions, despite their central role in achieving high performance on modern CPUs. As a result, vector-length agnosticism remains confined to execution rather than being reflected in the data representation, limiting the ability to apply layout-driven optimizations across the compilation pipeline. While this work establishes key compiler infrastructure for scalable vectors, its integration into full end-to-end compilation flows and its effectiveness across broader workloads remain only partially explored.

In the context of RVV, Peccia et al.~\cite{peccia2025tensor} integrate RVV support into TVM’s MetaSchedule framework, enabling efficient code generation for AI workloads through autotuned tensor programs. Their approach leverages RVV’s implementation-defined vector parameters and uses schedule search to select suitable vectorization strategies for a given implementation. While effective, this approach primarily addresses performance through implementation-specific tuning. In contrast, our work encodes vector-length agnosticism directly into the data layout, enabling representations that generalize across vector lengths without retuning.

While the above works primarily focus on compiler and code generation research, practical ML deployment today is often dominated by production frameworks and runtime systems. We therefore also consider widely used frameworks from the PyTorch~\cite{ansel2024pytorch} ecosystem. 

PyTorch's default eager execution mode executes operators directly through the runtime and dispatches them to backend implementations such as ATen (PyTorch’s tensor operator library) operators and underlying library kernels. While this model provides flexibility and broad operator coverage, it does not perform whole-graph or layout-aware optimizations.

TorchInductor, the default backend behind \texttt{torch.compile}, captures computation graphs and lowers them to a lower-level representation for code generation~\cite{ansel2024pytorch}. On CPUs, it relies on graph-level fusion, scheduling, and downstream compiler optimizations, but does not explicitly incorporate data-tiling abstractions tailored to scalable vector architectures.

ExecuTorch~\cite{executorch} follows a library-based approach by partitioning computation graphs and delegating supported subgraphs to optimized backend kernels such as XNNPack~\cite{xnnpack} and Arm-specific microkernels. This strategy can achieve strong performance for supported operator patterns, but is driven by subgraph matching and backend coverage rather than by introducing scalable data layouts within a general compilation pipeline.

Overall, these systems represent complementary design points: runtime execution (eager mode), graph compilation (Inductor), and delegated library execution (ExecuTorch). In contrast, our work focuses on end-to-end compiler code generation with scalable packed layouts, enabling data layouts and computations that adapt to implementation-defined vector length within a unified compilation pipeline.
\section{Methods and Implementation}
\label{sec:methods}

\subsection{Packed Matrix Multiplication}
\label{sec:packed_matmul}

Efficient matrix multiplication on CPUs relies on \emph{packed} (data-tiled) layouts, which reorganize operands in memory to match the access patterns of the underlying compute kernel.

Classical high-performance implementations (e.g., BLIS~\cite{BLIS1}-like designs) organize packing around cache-level tiles to maximize data reuse across the memory hierarchy, while ensuring that the microkernel operates on contiguous data. In this work, we adopt a similar packed formulation, but focus on layouts aligned with the register-level microkernel and treat packing as a standalone operation on the full operands of matrix multiplication, instead of packing operand tiles on the fly within the kernel execution loop.

Let $A \in \mathbb{R}^{M \times K}$ be a row-major matrix. In our setting, packing is defined with respect to the microkernel tile shape. A packed representation physically reorganizes $A$ into tiles of size $m_r \times k_r$ in memory, yielding a higher-rank tensor
\[
A^\text{pack} \in \mathbb{R}^{\lceil M/m_r \rceil \times \lceil K/k_r \rceil \times m_r \times k_r},
\]
with indexing relation
\[
A^\text{pack}[i_o, k_o, i_i, k_i] = A[i_o m_r + i_i,\; k_o k_r + k_i].
\]

Let $M_o = \lceil M / m_r \rceil$, $N_o = \lceil N / n_r \rceil$, and $K_o = \lceil K / k_r \rceil$ denote the number of tiles along each dimension.

Analogous tiling is applied to $B \in \mathbb{R}^{K \times N}$ using tile sizes $k_r$ and $n_r$, and to $C \in \mathbb{R}^{M \times N}$ using tile sizes $m_r$ and $n_r$.

The packed tensors are materialized in memory (typically in row-major order over their dimensions), such that elements within each tile are stored contiguously. Packing therefore corresponds to an explicit data transformation rather than a logical view.

Packed layouts are closely tied to the structure of the matrix multiplication kernel. For completeness, we illustrate a representative hierarchical loop structure that separates higher-level blocking (e.g., for cache or parallel execution) from the register-level packing used by the microkernel, as shown in Listing~\ref{lst:packed_gemm}.

\begin{lstlisting}[language=C, frame=single,
caption={Packed matrix multiplication loop structure},
  label={lst:packed_gemm}]
// Assume A_pack and B_pack are precomputed packed layouts
// T_M, T_N, and T_K denote higher-level blocking factors
// (e.g., for cache or parallel execution)
for (i_t = 0; i_t < M_o; i_t += T_M)
  for (j_t = 0; j_t < N_o; j_t += T_N)
    for (k_t = 0; k_t < K_o; k_t += T_K)

      for (i_o = 0; i_o < T_M; ++i_o)
        for (j_o = 0; j_o < T_N; ++j_o)
          for (k_o = 0; k_o < T_K; ++k_o)

            // Matmul microkernel on an m_r x n_r tile
            C_pack[i_t+i_o, j_t+j_o, :, :] +=
              A_pack[i_t+i_o, k_t+k_o, :, :] @
              B_pack[k_t+k_o, j_t+j_o, :, :]
\end{lstlisting}

The innermost computation corresponds to a \emph{microkernel}, which operates on small tiles held in registers and is typically unrolled to maximize instruction-level parallelism. The packed layouts are constructed such that each iteration of the microkernel consumes contiguous data, enabling efficient vectorization while avoiding register spilling.

Conventional packed layouts in high-performance BLAS-like libraries are primarily organized around cache-level blocking and are tightly coupled to specific microkernels, with tile sizes chosen to match the target architecture and kernel configuration. 

In a compiler setting, however, matrix multiplication is embedded within larger computation graphs, where operations may be fused and scheduled jointly. As a result, packing decisions must be exposed explicitly and defined at the register level to enable composability with surrounding operations and propagation across fused computations (see Sec.~\ref{sec:dt-formulation}).

In both cases, these layouts are typically defined using compile-time constants. While register-level tile sizes are often chosen based on the vector length of the target architecture, this assumption does not directly extend to scalable vector ISAs such as Arm SVE, where the vector length is not known at compile time.

\subsection{Scalable Packed Layouts}
\label{sec:scalable_packed_layouts}

We propose scalable packed layouts as an abstraction for representing data layouts parameterized by the hardware vector length, enabling portable and efficient code generation for scalable vector architectures.

Packed layouts are determined by the access pattern of the register-level microkernel that consumes them. In particular, the layout must ensure that the blocks of data loaded and stored by the microkernel are contiguous in memory, enabling efficient vector operations without additional data rearrangement during computation.

The structure of these access patterns is dictated by the instructions used within the microkernel together with its unrolling across different dimensions. For scalable vector ISAs, these instructions operate on vectors whose length is determined at runtime. As a result, the effective tile sizes of the microkernel, and therefore the granularity at which data must be laid out in memory, become functions of the hardware vector length $VL$ that is unknown at compile-time.

We adopt the packed tensor formulation introduced in Sec.~\ref{sec:packed_matmul}, where matrices are reorganized into register-level tiles of size $(m_r, n_r, k_r)$ that are materialized in memory prior to execution. In this formulation, the tile sizes are expressed as
\[
m_r = f_m(VL), \qquad
n_r = f_n(VL), \qquad
k_r = f_k(VL),
\]
where $VL$ denotes the hardware vector length in elements. The specific form of these functions is determined by the instruction-level access pattern and unrolling strategy of the microkernel.

\begin{figure}[t]
    \centering
    \includegraphics[width=\linewidth]{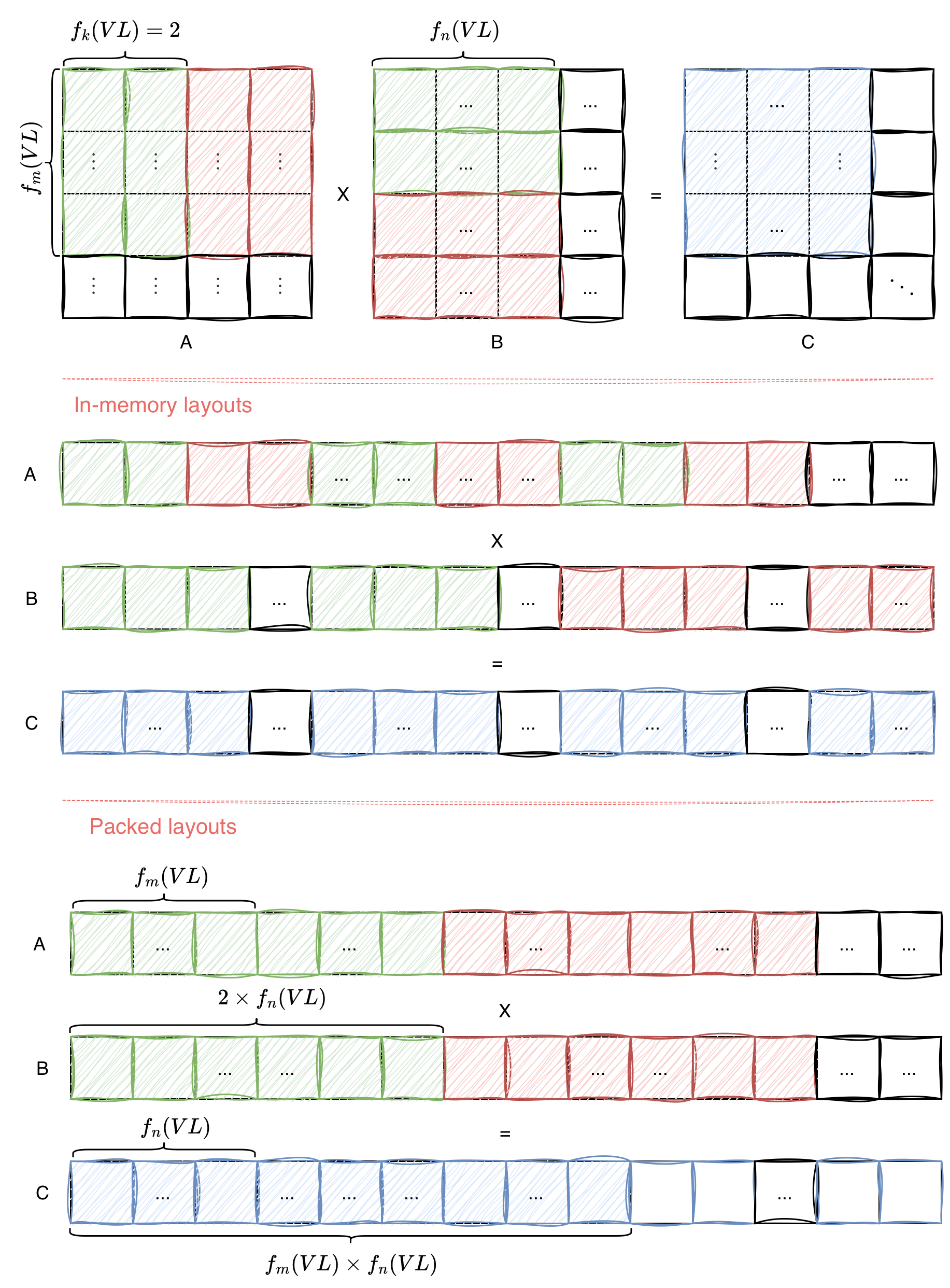}
    \caption{Representative transformation from a row-major layout to a scalable packed layout parameterized by the hardware vector length $VL$. The grouping of elements reflects the blocks consumed by the target microkernel and therefore scales with the hardware vector length.}
    \label{fig:scalable_packing}
\end{figure}

\paragraph{Scalable packing transformation.}
Fig.~\ref{fig:scalable_packing} illustrates the transformation from a row-major matrix to a scalable packed layout. The key property is that elements consumed together by the microkernel are grouped into contiguous regions in memory, matching the operand blocks required by the kernel. As a result, the layout cannot be represented by a fixed tiling.

\paragraph{Representative microkernel.}
To make this concrete, Listing~\ref{lst:scalable_microkernel_fp32} shows a representative FP32 microkernel targeting Arm SVE, which is also the kernel family used in our implementation. The kernel follows an outer-product formulation and operates on slices of the input matrices that are combined to update a register-resident tile of the output.

\begin{lstlisting}[language=C, frame=single,
caption={Representative intrinsic-level FP32 scalable-vector microkernel.},
label={lst:scalable_microkernel_fp32}]
void matmul_8x(2VL)xK_sve(float32_t *A, float32_t *B, float32_t *C) {
    size_t vlen = svcntw();

    svfloat32_t Acol0, Acol1;
    svfloat32_t Brow0, Brow1;
    svfloat32_t C0, C1, ..., C15;

    C0 = C1 = ... = C15 = svdup_n_f32(0);

    for (int i = 0; i < K; i++) {
        Acol0 = svld1rq_f32(svptrue_b32(), A + 8*i);
        Acol1 = svld1rq_f32(svptrue_b32(), A + 8*i + 4);

        Brow0 = svld1(svptrue_b32(), B + 2*vlen*i);
        Brow1 = svld1(svptrue_b32(), B + 2*vlen*i + vlen);

        C0 = svmla_lane_f32(C0, Brow0, Acol0, 0);
        C1 = svmla_lane_f32(C1, Brow1, Acol0, 0);
        C2 = svmla_lane_f32(C2, Brow0, Acol0, 1);
        C3 = svmla_lane_f32(C3, Brow1, Acol0, 1);
        /* ... */
        C14 = svmla_lane_f32(C14, Brow0, Acol1, 3);
        C15 = svmla_lane_f32(C15, Brow1, Acol1, 3);
    }

    svst1(svptrue_b32(), C, C0);
    /* ... */
    svst1(svptrue_b32(), C + 15*vlen, C15);
}
\end{lstlisting}

This kernel computes an outer-product update of an $8 \times 2VL$ tile of the output for each step along the $K$ dimension. At each iteration, it consumes an $8 \times 1$ slice from $A$ and a $1 \times 2VL$ slice from $B$. The former is loaded as two groups of four elements and replicated across vector registers, while the latter is loaded as two contiguous vectors of length $VL$.

The access pattern of this microkernel directly determines the required packed layout. In particular, the layout must provide contiguous slices matching exactly the operand blocks consumed by the kernel at each iteration.

\paragraph{Generality.}
The example above reflects a specific FP32 kernel on Arm SVE, but the underlying construction is more general. Different instruction sets, or data types induce different microkernel access patterns and therefore different functions $(f_m, f_n, f_k)$. The scalable packed layout formulation extends naturally to these cases by defining the layout in terms of the operand access pattern required by the target microkernel.

Importantly, this formulation does not restrict the design space of conventional packed layouts. Classical approaches explore different tiling strategies, kernel shapes, and tradeoffs between memory locality and compute efficiency; these choices are preserved in our formulation and expressed through the functions $(f_m, f_n, f_k)$. As a result, scalable packed layouts can represent the same range of packing strategies without restricting the design space.

In contrast to classical fixed-length SIMD designs, where $(m_r, n_r)$ are chosen as compile-time constants, our formulation expresses tile sizes as functions of the hardware vector length. This enables a single layout and kernel design to adapt across hardware implementations with different vector lengths. This formulation provides the foundation for generating scalable kernels and layouts within a compiler-driven workflow, as described in Sec.~\ref{sec:compiler}.

\subsection{Compiler Realization}
\label{sec:compiler}

\paragraph{Overview.}
This work extends existing data-tiled compilation in MLIR/IREE to support scalable packed layouts within the compiler pipeline. We integrate these layouts into key transformations, including tiling, fusion, vectorization, and bufferization, enabling their propagation throughout the compilation flow. Rather than relying on hand-written kernels and manually tuned packing strategies, the compiler identifies matrix multiplication operations in the input program and lowers them into a form that exposes register-level tiling and vectorization. The packed layout and corresponding microkernel are derived from a common specification, ensuring consistency between memory layout and computation. To support this, scalable layouts are represented using MLIR’s scalable vector abstractions, allowing them to be expressed symbolically and lowered to hardware-specific implementations during later compilation stages in MLIR, allowing the generated code to remain agnostic to the concrete hardware vector length while being lowered to hardware-specific instructions.

\paragraph{Data-tiled formulation.}
\label{sec:dt-formulation}
We build on the existing data-tiled formulation of matrix multiplication in the compiler IR, in which operations are decomposed into three stages: (i) packing of the input operands, (ii) computation on packed data, and (iii) unpacking of the result. This decomposition is represented using operations such as \texttt{linalg.pack}, \texttt{linalg.mmt4d}, and \texttt{linalg.unpack} of the MLIR \texttt{linalg} dialect~\cite{mlir}, where the central computation operates on already packed operands.

Our contribution extends this formulation to support scalable vector ISAs by generating packing and unpacking operations according to the scalable layouts defined in Sec.~\ref{sec:scalable_packed_layouts}. In particular, the tile sizes associated with these operations are expressed as functions of the hardware vector length $VL$ that is unknown at compile-time and the target microkernel structure. In contrast to classical BLAS-like approaches, which perform packing on-the-fly within the kernel and organize it around cache-level blocking, this formulation materializes packing as an explicit operation aligned with the register-level microkernel, enabling its composition and propagation across fused computations.

\paragraph{Kernel and layout generation.}
The compiler derives both the packed layouts and the corresponding compute kernels from a set of predefined layout configurations provided for the target hardware features and operand data types, selecting an appropriate microkernel structure (e.g., an outer-product formulation) and its associated layout for each matrix multiplication. These configurations define the tile sizes $(m_r, n_r, k_r)$ as functions of $VL$, which in turn guide both the packing transformations and the structure of the vectorized computation.

While the choice of microkernel and layout follows this predefined design space, the generation of the surrounding code is fully automatic. The resulting code closely resembles the representative microkernel shown in Listing~\ref{lst:scalable_microkernel_fp32}, while being generated automatically rather than written manually. The surrounding loop structure, including higher-level tiling and blocking, as well as fusion, vectorization, and bufferization, is determined by compiler heuristics.

\paragraph{Fusion and layout propagation.}
The explicit representation of packing and unpacking enables their placement to be optimized within the computation graph. The compiler partitions the program into subgraphs that can be efficiently lowered and schedules them according to heuristic cost models. Within this process, packing and unpacking operations can be fused into producers and consumers of the matrix multiplication, or propagated across adjacent operations when profitable, allowing the cost of packing to be amortized and, in many cases, hidden behind other computation.

As a result, scalable packed layouts are not confined to a single operation, but can influence the layout, tiling, and vectorization of surrounding computations. This propagation introduces additional constraints: operations that consume or produce packed data must respect the layout induced by $(m_r, n_r, k_r)$. To support this, we extend the tiling, fusion, and vectorization mechanisms in MLIR and IREE, modifying the corresponding passes to make them aware of scalable layouts during tile size selection, loop fusion, and vectorization. In particular, the compiler ensures that transformations applied to subgraphs remain consistent with the layout and vectorization strategy required by the target microkernel.

\paragraph{Padding semantics and vectorization.}
Packed layouts inherently incorporate padding semantics, as tiles are defined over rounded-up dimensions. When register-level tile sizes are aligned with these layouts, computations can be performed without additional masking, since out-of-bounds elements are represented explicitly in the packed data. This applies both to the core matrix multiplication operation and to other operations through which the packed layout is propagated, simplifying vectorization and improving efficiency. To enable the same effect for scalable vectors, we extend the \texttt{linalg} vectorization mechanisms to account for scalable vector types and tile sizes by providing additional information that allows the vectorizer to reason about the correspondence between tensor-level dynamic shapes and scalable vector representations, avoiding the introduction of unnecessary masking.

\paragraph{Extensibility.}
The presented mechanism enables the use of scalable packed layouts within a production-grade compiler. While demonstrated here for FP32 matrix multiplication on Arm SVE, the approach generalizes to other relevant data types and scalable vector ISAs. As long as the lowering from vector operations to target instructions is supported, the compiler can generate corresponding kernels and layouts. This makes the approach extensible while preserving the benefits of automated code generation.
\section{Results and Evaluation}
\label{sec:results}
We evaluate the performance of code generated by our approach on two consumer-grade Arm-based devices: a Radxa Orion O6 board and a Google Pixel~9. Tab.~\ref{tab:specs} summarizes the hardware specifications of both platforms. All results are generated with our open-source implementation of IREE (SVE) code generation, which is in the process of being upstreamed to the mainline IREE and LLVM projects~\cite{UPSTEAMLINK}. All benchmarks reported in Sec.~\ref{sec:neon_exp} and Sec.~\ref{sec:torch_exp} are executed $50$ times with additional warm-up iterations, and we report average latencies.

\begin{table}
  \caption{Hardware Specifications of the Radxa Orion O6 and Google Pixel 9}
  \label{tab:specs}
  \begin{tabular}{lll}
    \toprule
    Spec. & \textbf{Orion O6 (CIX P1)} & \textbf{Pixel 9 (Tensor G4)} \\
    \midrule
    Process       & TSMC 6\,nm              & Samsung 4\,nm            \\
    CPU Cores     & 12                       & 8                        \\
    Big           & 4$\times$A720 @2.6\,GHz & 1$\times$X4 @3.1\,GHz   \\
    Mid           & 4$\times$A720 @2.4\,GHz & 3$\times$A720 @2.6\,GHz \\
    Little        & 4$\times$A520 @1.8\,GHz & 4$\times$A520 @1.92\,GHz\\
    Memory        & 64\,GB LPDDR5           & 12\,GB LPDDR5X           \\
    SVE Length    & 128\,bit                & 128\,bit                 \\
    OS            & Ubuntu 24.04.4 LTS      & Android 16 (API 36)      \\
    \bottomrule
  \end{tabular}
\end{table}

In Tab.~\ref{tab:model-index} one can see all models used in our evaluation, which are publicly available on HuggingFace and are cast to FP32 precision prior to benchmarking if not already in that format. All reported latency numbers correspond to a single forward pass. Since PyTorch models accept \textit{dynamic} input sizes between forward passes, we fix a consistent input shape per model across all frameworks during export, compilation, and benchmarking. We set the batch size to~$1$ for all models, reflecting the single-request inference pattern typical of consumer-grade deployments.

\begin{table}[t]
  \centering
  \small
  \caption{Model indices used in the IREE (SVE) speedup figures.
           Indices~1--9 appear on both Orion~O6 and Pixel 9;
           indices~10--11 are Orion~O6 only.}
  \label{tab:model-index}
  \begin{tabular}{@{}rl@{}}
    \toprule
    Model Index & \textbf{HuggingFace Model} \\
    \midrule
     1 & \texttt{HuggingFaceTB/SmolLM2-135M} \\
     2 & \texttt{HuggingFaceTB/SmolVLM-256M-Instruct} \\
     3 & \texttt{google/mobilebert-uncased} \\
     4 & \texttt{google/vit-base-patch16-224} \\
     5 & \texttt{hustvl/yolos-tiny} \\
     6 & \texttt{Qwen/Qwen2.5-Coder-0.5B} \\
     7 & \texttt{MIT/ast-finetuned-audioset-10-10-0.4593} \\
     8 & \texttt{openai/whisper-base} \\
     9 & \texttt{FacebookAI/xlm-roberta-large} \\
    10 & \texttt{deepseek-ai/DeepSeek-R1-Distill-Qwen-1.5B} \\
    11 & \texttt{meta-llama/Llama-3.2-3B-Instruct} \\
    \bottomrule
  \end{tabular}
\end{table}

\begin{figure*}[htb]
    \centering
    \begin{subfigure}[t]{0.31\textwidth}
        \centering
        \includegraphics[width=\linewidth]{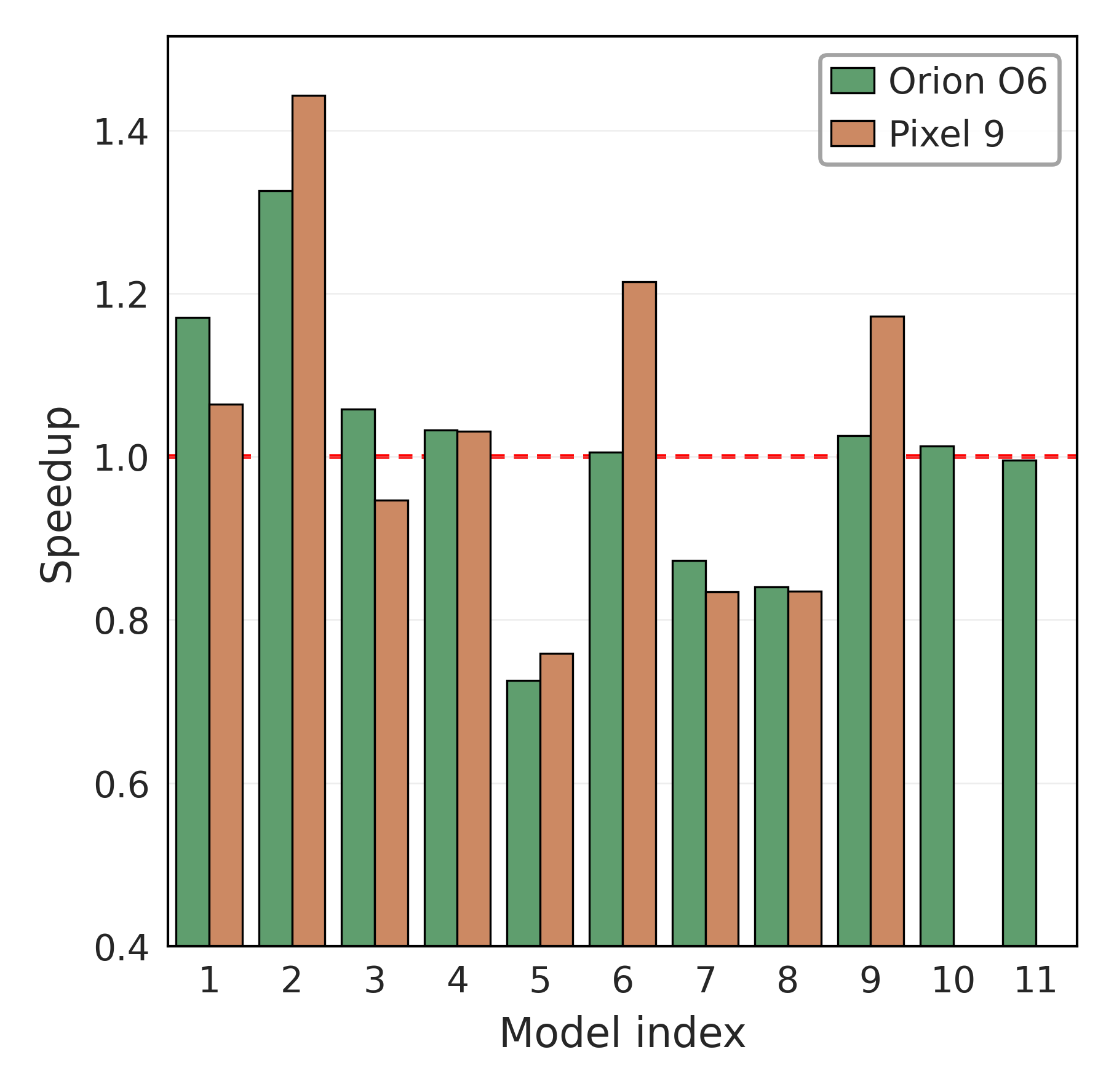}
        \caption{IREE (SVE) vs IREE (NEON)}
        \label{fig:sub_sve_neon}
    \end{subfigure}
    \hfill
    \begin{subfigure}[t]{0.31\textwidth}
        \centering
        \includegraphics[width=\linewidth]{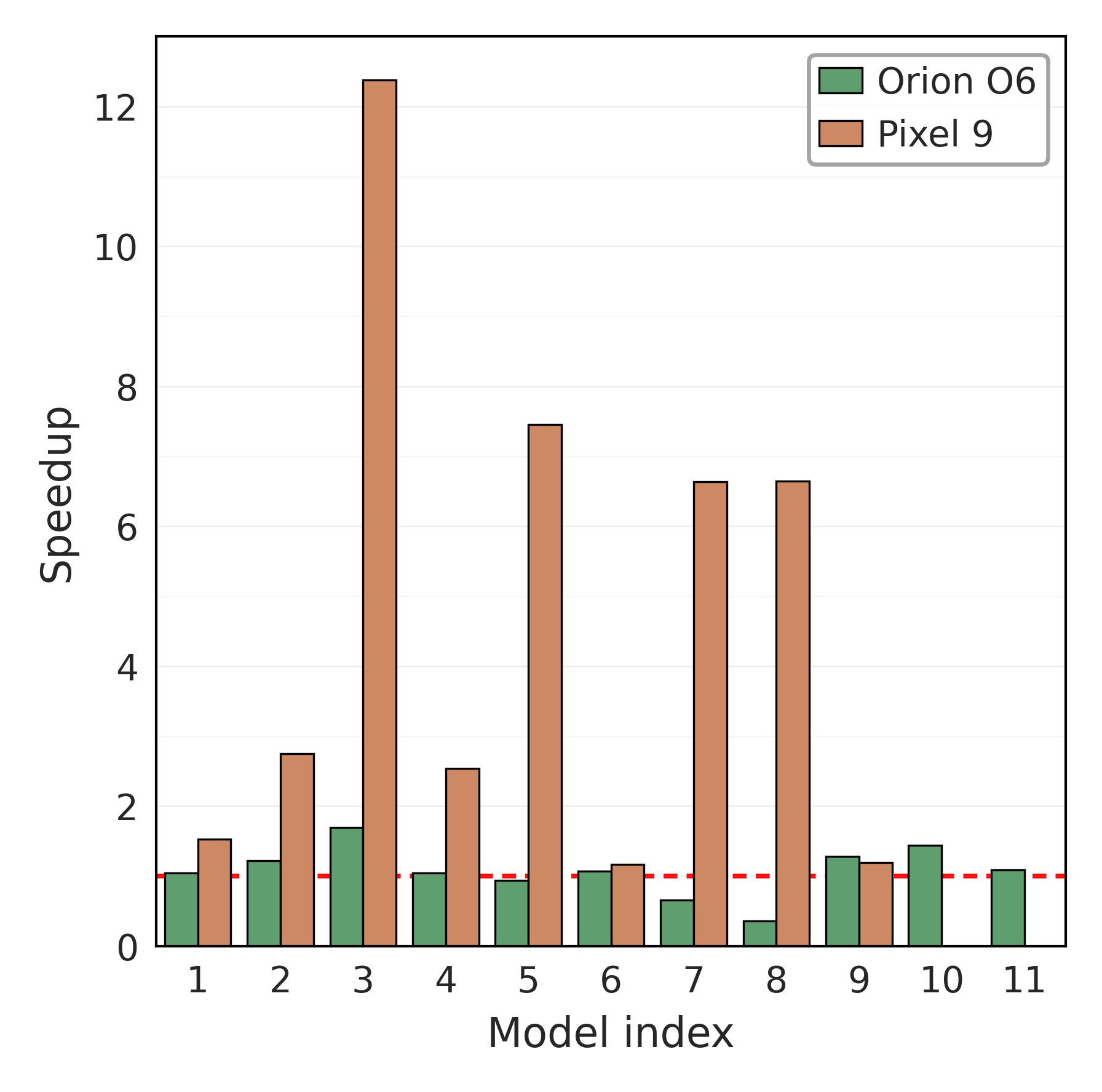}
        \caption{IREE (SVE) vs ExecuTorch}
        \label{fig:sub_sve_et}
    \end{subfigure}
    \hfill
    \begin{subfigure}[t]{0.31\textwidth}
        \centering
        \includegraphics[width=\linewidth]{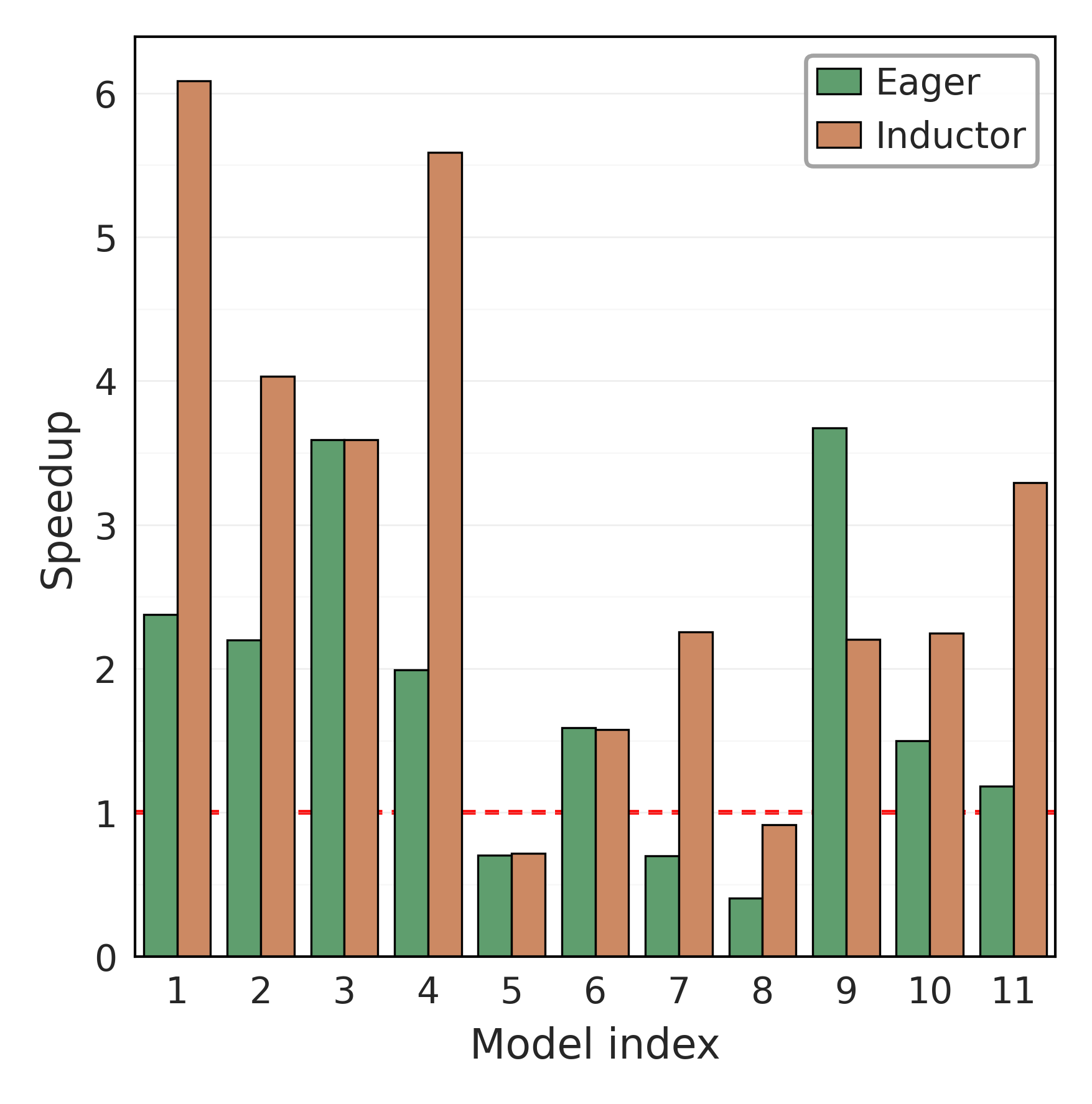}
        \caption{IREE (SVE) vs Inductor/Eager O6}
        \label{fig:sub_sve_inductor_eager}
    \end{subfigure}
    \caption{Speedups achieved with our IREE (SVE) code generation approach against (\ref{fig:sub_sve_neon}) the existing NEON pipeline in IREE, (\ref{fig:sub_sve_et}) ExecuTorch, and (\ref{fig:sub_sve_inductor_eager}) TorchInductor and eager mode (only on Orion O6). Model indices are to be taken from Tab.~\ref{tab:model-index}.}
\end{figure*}

\subsection{Comparison to NEON Codegen}
\label{sec:neon_exp}

First, we compare our scalable IREE (SVE) code generation approach against the existing static NEON code generation in IREE. This comparison isolates the impact of VLA code generation within the same compiler stack, allowing us to assess whether the additional flexibility of scalable vectors introduces performance overheads compared to fixed-length SIMD. Speedups achieved with multi-threaded execution are shown in Fig.~\ref{fig:sub_sve_neon} and Tab.~\ref{tab:sve-vs-neon}.

\begin{table}[t]
  \centering
  \caption{Our IREE (SVE) vs.\ IREE (NEON) latency (ms) using 12 threads on Orion O6 and 8 threads on Pixel 9. Percentage in parentheses shows IREE (NEON) overhead relative to IREE (SVE); negative values indicate that IREE (NEON) is faster. Model names are abbreviated but correspond to those from Tab.~\ref{tab:model-index}.}
  \label{tab:sve-vs-neon}
  \small
  \begin{tabular}{@{}l rr rr@{}}
  \toprule
  & \multicolumn{2}{c}{\textbf{Radxa Orion O6}} & \multicolumn{2}{c}{\textbf{Google Pixel 9}} \\
  \cmidrule(lr){2-3} \cmidrule(lr){4-5}
  \textbf{Model}
    & \shortstack{SVE\\(ms)}
    & \shortstack{NEON\\(ms)}
    & \shortstack{SVE\\(ms)}
    & \shortstack{NEON\\(ms)} \\
  \midrule
  DeepSeek-R1-1.5B   & 3470 & 3515\,(+1\%)   & \multicolumn{2}{c}{---}           \\
  Llama-3.2-3B       & 4978 & 4957\,($-$0\%) & \multicolumn{2}{c}{---}           \\
  Qwen2.5-0.5B       & 782  & 786\,(+1\%)    & 1802 & 2188\,(+21\%)  \\
  SmolLM2-135M       & 246  & 288\,(+17\%)   & 730  & 777\,(+6\%)    \\
  SmolVLM-256M       & 319  & 423\,(+33\%)   & 906  & 1307\,(+44\%)  \\
  AST                & 2571 & 2243\,($-$13\%) & 4871 & 4063\,($-$17\%) \\
  MobileBERT         & 79   & 84\,(+6\%)     & 206  & 195\,($-$5\%)  \\
  ViT-Base           & 185  & 191\,(+3\%)    & 552  & 569\,(+3\%)    \\
  Whisper-Base       & 1811 & 1521\,($-$16\%) & 2571 & 2146\,($-$17\%) \\
  XLM-RoBERTa-Large  & 630  & 646\,(+3\%)    & 2241 & 2627\,(+17\%)  \\
  YOLOS-Tiny         & 623  & 452\,($-$27\%) & 1306 & 991\,($-$24\%) \\
  \bottomrule
  \end{tabular}
\end{table}


Our approach outperforms IREE (NEON) code in 7 out of 11 models on the Orion O6 board, achieving up to $1.33\times$ speedup, and in 5 out of 9 models on the Google Pixel~9, achieving up to $1.45\times$ speedup. The two models absent from the Pixel~9 results could not be benchmarked due to memory constraints.

Both devices implement SVE and NEON with the same vector length of 128~bits, so performance differences cannot be attributed to vector length. Instead, they arise from differences in code generation strategy and data layout, as the static (NEON) and scalable (SVE) vector abstractions expose different optimization opportunities to the compiler, leading to measurable variation. On the one hand, scalable vector types are opaque at compile time, which can prevent the compiler from inferring memory alignment and lead to conservative predicate masks on loads and stores, incurring penalties of up to 27\% in our measurements. On the other hand, we attribute the speedups of our approach to two factors: (1)~SVE's first-class predicate registers and dedicated loop-control instructions (e.g., \texttt{WHILELT}) enable efficient tail handling and eliminate the scalar cleanup loops and branch-heavy epilogues typical of NEON code, thereby reducing both instruction count and branch misprediction overhead; and (2)~the compiler's cost model makes different tiling and scheduling decisions when the vector length is unknown at compile time, which can yield more favorable decompositions for certain workloads, particularly in multi-threaded settings.

Overall, the vector length-agnostic programming paradigm introduces additional challenges for code generation, and these results demonstrate that our approach meets them effectively, producing code that is competitive with its NEON counterpart. Moreover, because our generated code is vector-length-agnostic and compatible with any vector length permitted by the SVE specification, the same binaries can transparently adapt to wider SIMD hardware without recompilation, enabling performance portability across implementations with different vector lengths. NEON, by contrast, is fixed at 128~bits and requires retuning or redesign of kernels to effectively exploit wider SIMD units on future architectures. We verify this portability property quantitatively on a controlled platform in Sec.~\ref{subsec:gem5-sim}.

\subsection{Comparison to PyTorch Ecosystem Frameworks}
\label{sec:torch_exp}

We compare our SVE code generation in IREE against ExecuTorch (v1.1.0), TorchInductor, and PyTorch eager execution, both from PyTorch v2.10.0, as these widely used frameworks are representative of modern deployment and execution strategies for PyTorch models in both industry and academia. For ExecuTorch, we use its standard XNNPack~\cite{xnnpack} backend to ensure comparison against optimized production kernels. XNNPack dispatches to KleidiAI~\cite{kleidiai} microkernels for AArch64. At the time of writing, KleidiAI provides NEON, SVE, and SME microkernels for FP32 element types, with the specific implementation selected at runtime based on device capabilities. Since neither of our target devices supports the Arm SME extension, subgraphs delegated to XNNPack are restricted to NEON and SVE microkernels. TorchInductor and PyTorch's eager execution mode are evaluated only on the Orion O6, as neither is available on Android due to the lack of native Python support on the Pixel~9.
\paragraph{ExecuTorch.}
Speedups achieved with our approach are shown in Fig.~\ref{fig:sub_sve_et} and Tab.~\ref{tab:et-vs-iree}.
\begin{table}[t]
  \centering
  \caption{ExecuTorch latency relative to our IREE (SVE) implementation. Thread counts in parentheses denote the configuration used by ExecuTorch to achieve minimum latency. Our approach consistently achieves optimal performance at the maximum thread counts ($12$ for the Orion O6 and $8$ for the Pixel 9). A ratio $> 1$ indicates that IREE is faster.}
  \label{tab:et-vs-iree}
  \small
  \begin{tabular}{@{}l rr rr@{}}
  \toprule
  & \multicolumn{2}{c}{\textbf{Radxa Orion O6}} & \multicolumn{2}{c}{\textbf{Google Pixel 9}} \\
  \cmidrule(lr){2-3} \cmidrule(lr){4-5}
  \textbf{Model}
    & \shortstack{ExecuTorch\\(ms)}
    & {ET\,/\,IREE}
    & \shortstack{ExecuTorch\\(ms)}
    & {ET\,/\,IREE} \\
  \midrule
  DeepSeek-R1-1.5B   & 5000\,(8)   & 1.44$\times$ & 14469\,(4)  & ---            \\
  Llama-3.2-3B       & 5400\,(8)   & 1.08$\times$ & \multicolumn{2}{c}{---}        \\
  Qwen2.5-0.5B       & 835\,(8)    & 1.07$\times$ & 2109\,(4)   & 1.17$\times$   \\
  SmolLM2-135M       & 256\,(8)    & 1.04$\times$ & 1119\,(4)   & 1.53$\times$   \\
  SmolVLM-256M       & 388\,(12)   & 1.22$\times$ & 2495\,(4)   & 2.75$\times$   \\
  AST                & 1700\,(8)   & 0.66$\times$ & 32324\,(8)  & 6.64$\times$   \\
  MobileBERT         & 135\,(4)    & 1.70$\times$ & 2550\,(4)   & 12.38$\times$  \\
  ViT-Base           & 194\,(12)   & 1.05$\times$ & 1404\,(8)   & 2.54$\times$   \\
  Whisper-Base       & 647\,(12)   & 0.36$\times$ & 17084\,(8)  & 6.64$\times$   \\
  XLM-RoBERTa-Large  & 806\,(12)   & 1.28$\times$ & 2678\,(4)   & 1.20$\times$   \\
  YOLOS-Tiny         & 585\,(8)    & 0.94$\times$ & 9735\,(8)   & 7.45$\times$   \\
  \bottomrule
  \end{tabular}
\end{table}

Our approach outperforms ExecuTorch on most evaluated models on the Orion O6, achieving lower latency in 8 of 11 cases and up to $1.70\times$ speedup, and in 9 out of 9 models on the Google Pixel~9, achieving up to $12.38\times$ speedup. The largest gains occur on matrix multiplication-dominated models, particularly transformer workloads with substantial attention computation. This is consistent with the focus of our work on SVE-aware data-tiling, which primarily targets matrix multiplication kernels.

In contrast, models with convolution operators or convolution-heavy front-ends, such as \texttt{whisper-base}, exhibit higher latency under our approach. This behavior is consistent with the current lack of direct data-tiling support for convolutions in both our system and baseline IREE on CPUs. Instead, convolutions are lowered to matrix multiplication--like forms via transformations such as \emph{im2col}~\cite{chellapilla2006high}. This lowering can explicitly materialize large intermediate matrices containing duplicated input patches, increasing memory footprint and memory traffic, and can therefore be less efficient than the specialized convolution kernels used by ExecuTorch. Supporting direct data-tiling for convolution operators is a natural direction for future work. 

\paragraph{TorchInductor and PyTorch Eager.}
Speedups achieved with our approach are shown in Fig.~\ref{fig:sub_sve_inductor_eager} and Tab.~\ref{tab:e2e-eager-inductor}. 
\begin{table}[t]
  \centering
  \caption{PyTorch eager execution mode and Inductor latency on Radxa Orion O6 relative to IREE (SVE). Thread counts in parentheses. Ratio $> 1$ indicates IREE is faster.}
  \label{tab:e2e-eager-inductor}
  \small
  \begin{tabular}{@{}l rr rr@{}}
  \toprule
  \textbf{Model}
    & \shortstack{Eager\\(ms)}
    & {Eager\,/\,IREE}
    & \shortstack{Inductor\\(ms)}
    & {Ind.\,/\,IREE} \\
  \midrule
  DeepSeek-R1-1.5B   & 5200\,(12)  & 1.50$\times$ & 7800\,(8)   & 2.25$\times$ \\
  Llama-3.2-3B       & 5900\,(12)  & 1.19$\times$ & 16400\,(12) & 3.29$\times$ \\
  Qwen2.5-0.5B       & 1244\,(12)  & 1.59$\times$ & 1234\,(12)  & 1.58$\times$ \\
  SmolLM2-135M       & 584\,(12)   & 2.37$\times$ & 1497\,(1)   & 6.09$\times$ \\
  SmolVLM-256M       & 701\,(12)   & 2.20$\times$ & 1286\,(4)   & 4.03$\times$ \\
  AST                & 1800\,(12)  & 0.70$\times$ & 5800\,(12)  & 2.26$\times$ \\
  MobileBERT         & 285\,(8)    & 3.59$\times$ & 284\,(8)    & 3.59$\times$ \\
  ViT-Base           & 369\,(12)   & 1.99$\times$ & 1034\,(1)   & 5.59$\times$ \\
  Whisper-Base       & 736\,(12)   & 0.41$\times$ & 1657\,(12)  & 0.92$\times$ \\
  XLM-RoBERTa-Large  & 2313\,(12)  & 3.67$\times$ & 1389\,(12)  & 2.20$\times$ \\
  YOLOS-Tiny         & 439\,(12)   & 0.70$\times$ & 446\,(12)   & 0.72$\times$ \\
  \bottomrule
  \end{tabular}
\end{table}
Our approach also outperforms PyTorch's eager execution mode in 8 of 11 models, with speedups of up to $3.67\times$, and TorchInductor in 9 of 11 models, with speedups of up to $6.09\times$. PyTorch eager execution mode runs models without prior graph capture or whole-graph optimization, evaluating operators immediately as they are encountered and dispatching them through the runtime to backend implementations. In this mode, every line of Python is executed in sequence, while TorchInductor changes execution by capturing and compiling graphs. As a result, eager execution cannot exploit fusion, global scheduling, or layout-aware transformations across operator boundaries in the same way as compiler-based approaches, which likely explains why our approach outperforms eager execution on most models.

TorchInductor mitigates some of these limitations by capturing and compiling graphs, and it is the default compiler backend behind \texttt{torch.compile}. However, our results suggest that its current CPU backend is not well optimized for our target AArch64 processors. In particular, when our approach wins, the gap to TorchInductor is often substantially larger than the gap to eager execution, suggesting that the generated code does not consistently exploit the architectural features of these CPUs as effectively as either our approach or baseline IREE. At the same time, this observation should not be overgeneralized: PyTorch has reported strong TorchInductor results on AWS Graviton3 processors after substantial backend tuning and reuse of Arm-specific optimized kernels~\cite{pytorch_aws}. 
We therefore interpret our results as evidence that TorchInductor's current CPU backend does not yet generalize uniformly across diverse Arm platforms.

\paragraph{Compiler vs. library approaches.}
More broadly, library-based systems such as ExecuTorch rely on subgraph matching to dispatch recognized operator patterns to hand-optimized kernels. This strategy can deliver excellent performance for supported patterns, especially for operators such as convolutions, but it may miss optimization opportunities that span larger regions of the graph.

Compiler-based approaches such as ours can instead optimize across operator boundaries and apply target-aware whole-graph transformations. In addition, our approach introduces scalable tensor layouts, enabling consistent code generation without relying on per-target kernel implementations. We therefore view the two paradigms as complementary: library-based systems provide strong out-of-the-box performance for well-supported operator patterns, while compiler-based approaches offer greater flexibility and broader opportunities for global optimization.
\paragraph{Limitations of current hardware.}
A central motivation for scalable vectorization is performance portability across implementations with different vector lengths. However, this property is difficult to isolate empirically on currently accessible hardware. Both platforms used in this study implement SVE with a vector length of 128 bits, matching NEON, and therefore do not expose the primary advantage of scalable vectors. Even on platforms with wider SVE support, empirical scaling remains difficult to observe. For example, AWS Graviton~3 implements a 256-bit SVE vector length, but provides fewer SVE execution units than NEON units, resulting in comparable throughput between the two execution modes. This illustrates that vector length alone does not determine performance, and that microarchitectural factors such as the number of execution units can dominate.

Beyond standard SVE, some recent processors support the Arm SME, which introduces a streaming mode with a potentially larger streaming vector length (SVL) than the base SVE vector length. In principle, this mode could be used to evaluate scaling behavior by executing vector-vector operations on wider registers without relying on matrix-specific instructions.  However, our experiments on a Qualcomm Snapdragon~8 Elite Gen~5 device (featuring a 512-bit SVL), as well as prior observations from Remke and Breuer~\cite{remke2024hello} on Apple M4-class processors, show no clear performance scaling for such workloads when implemented using vector (non-matrix) instructions, and in some cases even reduced throughput. We note that SME provides dedicated matrix units that can achieve higher performance for matrix multiplication workloads; however, these results focus specifically on vector-based implementations to isolate the effect of increased vector length. These observations suggest that current implementations may not translate increased vector length into proportional performance gains for general vector workloads.

Finally, while large-scale systems such as the Fugaku supercomputer~\cite{fugaku_system} provide wider SVE implementations, they are not readily accessible for experimentation. To characterise the vector-length scaling behaviour of our generated code despite the lack of suitable real-hardware platforms, we complement our on-device evaluation with a controlled simulator-based study, presented in Sec.~\ref{subsec:gem5-sim}.

\subsection{Simulator-Based Scaling Study}
\label{subsec:gem5-sim}

To isolate vector-length scaling in a controlled setting, we run our compiled binaries through the gem5 cycle-level simulator~\cite{binkert2011gem5, lowe2020gem5}, building on Brank's~\cite{brank2023vector} Neoverse-N1 micro-architecture model. The three simulated variants (SVE-128, SVE-256, SVE-512) differ \emph{only} in their SVE horizontal-reduction latencies, scaled per vector length; all other pipeline, cache, and branch-prediction parameters are identical, so any cycle difference is attributable to the vector length alone. The simulated core is a single-threaded 8-wide out-of-order AArch64 pipeline at 2.5~GHz with 64~KiB L1 caches, a 1~MiB L2 with a tagged hardware prefetcher enabled, and a shared L3 evaluated at 8~MiB and 16~MiB, covering the range commonly found in consumer-grade Arm SoCs. Main memory is a four-channel LPDDR5-6400 configuration with $\sim$51~GB/s peak bandwidth.

\begin{figure}[t]
    \centering
    \includegraphics[width=0.8\linewidth]{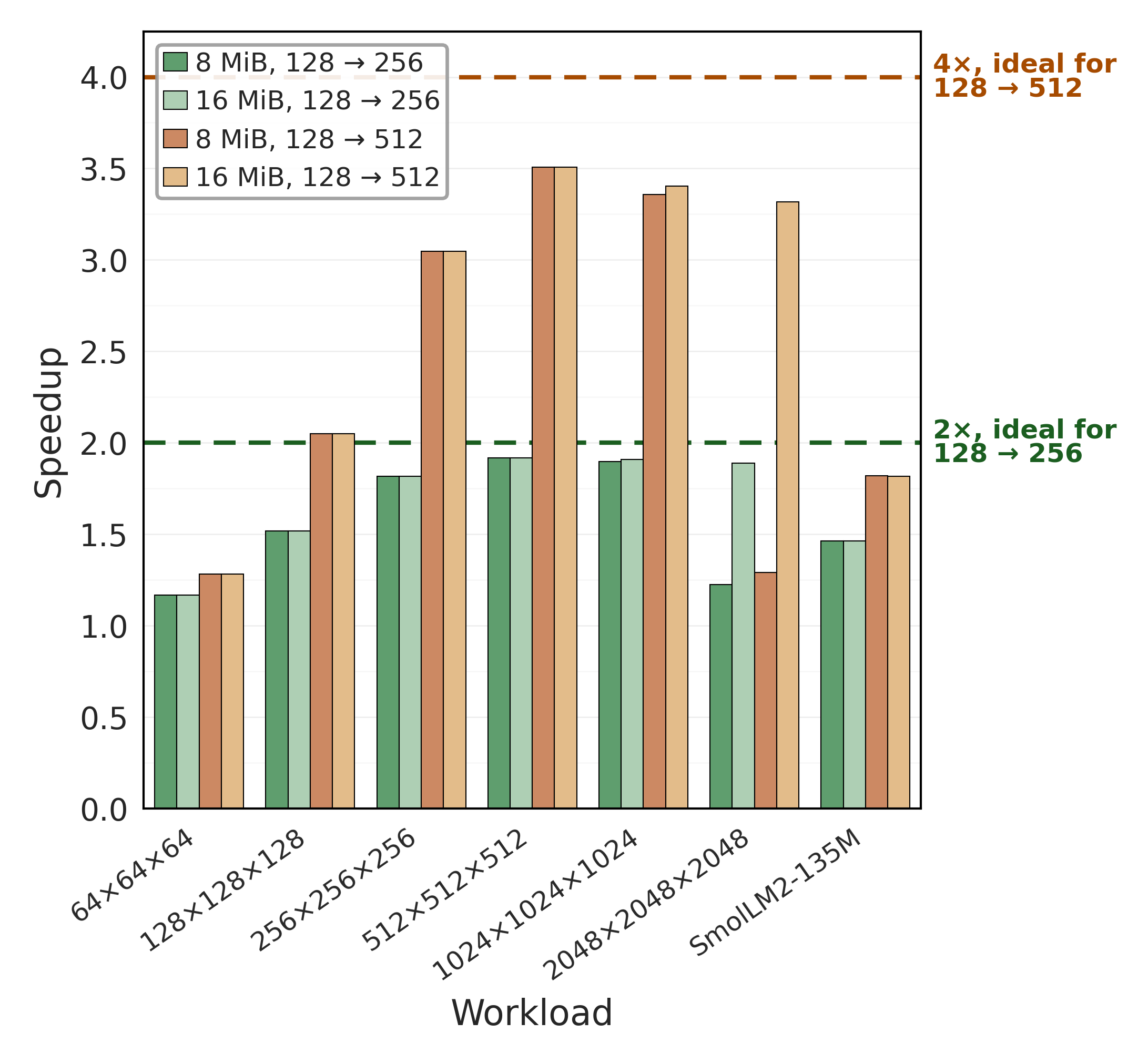}
    \caption{Speedup of our scalable SVE code generation relative to SVE-128 when the vector length is widened to 256 and 512~bits, for square FP32 matmuls of varying size, for the skinny-K variant, and for an end-to-end forward pass of \texttt{SmolLM2-135M}.}
    \label{fig:sve-width-scaling}
\end{figure}

Fig.~\ref{fig:sve-width-scaling} reports the SVE-256 and SVE-512 speedups relative to SVE-128 for square FP32 matmuls at $N \in \{64, 128, 256, 512, 1024, 2048\}$ and a skinny-K variant ($2048 \times 2048 \times 512$) representative of LLM-style aspect ratios.

The cache-resident matmuls scale nearly ideally: the $1024\times1024\times1024$ case reaches $3.36\times$ at SVE-512 and the skinny-K variant reaches $3.44\times$. The remaining gap to the ideal $4\times$ ceiling is attributable to the higher horizontal-reduction latencies at wider widths. For these sizes ($N \le 1024$ and skinny-K), the 8~MiB and 16~MiB L3 configurations produce near-identical cycle counts.

At $N=2048$, the compute-bound to memory-bound transition becomes visible. The three $2048 \times 2048$ FP32 operands total $48$~MiB, $6\times$ the 8~MiB L3, so tiles loaded into cache are evicted before they can be reused across outer-loop iterations: operands are refetched heavily from DRAM, the three widths converge on similar cycle counts, and the SVE-512 speedup collapses to $1.29\times$. With a 16~MiB L3, the reused tiles stay resident between passes: DRAM traffic drops by roughly an order of magnitude, and the scaling recovers to $3.32\times$ at SVE-512.

\paragraph{End-to-end inference on SmolLM2-135M.}
We additionally evaluate an end-to-end forward pass of \texttt{SmolLM2-135M} at sequence length~32. The forward pass achieves $1.46\times$ and $1.82\times$ speedups at SVE-256 and SVE-512 respectively. Although the model is matmul-dominated, the surrounding operations (attention softmax, layer normalisation, element-wise non-linearities) are not covered by our scalable packed layouts and are generally memory-bound, so they do not scale linearly with the vector length and cap the overall end-to-end speedup.

Overall, our vector-length-agnostic code generation delivers near-ideal vector length scaling on compute-bound matmul kernels, and translates into meaningful end-to-end scaling on a realistic transformer workload, the remaining gap being dictated by memory-bound non-matmul operations rather than limitations of the generated code.

\section{Conclusion}
\label{sec:conclusion}
We presented an approach for enabling vector-length-agnostic (VLA) code generation in an end-to-end machine learning compilation pipeline through vector-length-parametric packed data layouts and compiler extensions. We integrated these methods into a production-grade MLIR/IREE-based pipeline, extending tiling, fusion, and vectorization.

Our evaluation on Arm SVE shows that the added complexity of VLA code generation can be handled effectively. The generated code is competitive with, and often outperforms, existing NEON-based implementations within IREE and state-of-the-art PyTorch frameworks. A controlled simulator study further shows that performance scales as expected with increasing SVE vector length on compute-bound workloads, supporting portability where hardware is not yet available. While we focus on FP32 matrix multiplication, further optimization opportunities remain in microkernel design and tuning, and in extending support to additional data types.

A limitation of the current approach is that the packed layout and transformations are tailored to matrix multiplication. Although these layouts can propagate through parts of the computation graph, they do not directly support convolutions. These are lowered via \texttt{im2col} to reuse the same packing strategy, incurring overhead and limiting efficiency. Addressing this requires dedicated data layouts, lowering strategies, and specialized microkernels.

Looking forward, we are extending this work in several directions. We are exploring lower-precision arithmetic (including BF16) and quantized models for edge deployment, as well as extending the approach to convolutions with dedicated layouts and lowering strategies beyond \texttt{im2col}. In addition, we are targeting other scalable vector architectures, including RVV and emerging extensions such as Arm SME.

Overall, this work demonstrates that scalable vector architectures can be effectively integrated into modern ML compiler pipelines, enabling more portable execution of machine learning workloads.
\bibliographystyle{acm}
\bibliography{references}
\end{document}